\documentclass[sigconf]{acmart}

\usepackage{balance}

\def\BibTeX{{\rm B\kern-.05em{\sc i\kern-.025em b}\kern-.08emT\kern-.1667em\lower.7ex\hbox{E}\kern-.125emX}}
    
\usepackage{lipsum}
\usepackage{xspace}

\usepackage{algorithm}
\usepackage{algorithmicx, algpseudocode}

\newcommand{\mmpfc}{IBiS\xspace}
\newcommand{\mmrpfc}{IBiS$^{RP}$\xspace}
\newcommand{\mmrpfcdac}{IBiS$^{RP+DAC}$\xspace}
\newcommand{\mmrpfcdacvls}{IBiS$^{RP+DAC-VLS}$\xspace}
\newcommand{\mmrpfcdacvlsb}{IBiS$^{RP+DAC+DAC-VLS}$\xspace}
\newcommand{\vllcp}{llcp-only\xspace}
\newcommand{\vrlcp}{rlcp-only\xspace}
\newcommand{\vsinceros}{\ensuremath{no}-\ensuremath{term}\xspace}

\usepackage{ifthen}

\newcommand{\lookup}[1][] {\ifthenelse{ \equal{#1}{} }
{\emph{lookup}\xspace}
{\emph{lookup(#1)}\xspace}
}
      
\newcommand{\access}[1] [] {\ifthenelse{ \equal{#1}{} }
      {\emph{access}\xspace}
      {\emph{access(#1)}\xspace}
}

\newcommand{\psubl}{\ensuremath{p_\ell}\xspace}
\newcommand{\psubr}{\ensuremath{p_r}\xspace}
\newcommand{\psubm}{\ensuremath{p_m}\xspace}
\newcommand{\psubi}{\ensuremath{p_i}\xspace}
\newcommand{\lcp}{\ensuremath{lcp}\xspace}
\newcommand{\llcp}{\ensuremath{llcp}\xspace}
\newcommand{\rlcp}{\ensuremath{rlcp}\xspace}
\newcommand{\maxlcp}{\ensuremath{maxlcp}\xspace}
\newcommand{\xll}{\ensuremath{\ell}\xspace}
\newcommand{\xrr}{\ensuremath{r}\xspace}
\newcommand{\sq}{\ensuremath{s_q}\xspace}
\newcommand{\off}{\ensuremath{o}\xspace}

\newcommand{\duk}{UK\xspace}
\newcommand{\darabic}{Arabic\xspace}
\newcommand{\duris}{URIs\xspace}
\newcommand{\dliterals}{Literals\xspace}

\newcommand{\pfc}{PFC\xspace}
\newcommand{\rpfc}{RPFC\xspace}

\newcommand{\rphtfc}{RPHTFC\xspace}

\newcommand{\hashrp}{HASHRPDAC\xspace}
\newcommand{\rpdac}{RPDAC\xspace}
\newcommand{\pdt}{PDT\xspace}

\setcopyright{acmlicensed}

\begin{document}

\copyrightyear{2019}
\acmYear{2019}
\acmConference[CIKM '19]{The 28th ACM International Conference on Information and Knowledge Management}{November 3--7, 2019}{Beijing, China}
\acmBooktitle{The 28th ACM International Conference on Information and Knowledge Management (CIKM '19), November 3--7, 2019, Beijing, China}
\acmPrice{15.00}
\acmDOI{10.1145/3357384.3357972}
\acmISBN{978-1-4503-6976-3/19/11}

\fancyhead{}

%
\title{Improved Compressed String Dictionaries}

%
\author{Nieves R. Brisaboa}
\email{brisaboa@udc.es}
\orcid{0000-0001-8025-3048}
\affiliation{%
  \institution{Universidade da Coru\~na, \\ Centro de investigaci\'on CITIC, Databases Lab.}
  \streetaddress{Campus de Elvi\~na s/n}
  \city{A Coru\~na}
  \state{Spain}
  \postcode{15071}
}

\author{Ana Cerdeira-Pena}
\email{acerdeira@udc.es}

\affiliation{%
  \institution{Universidade da Coru\~na, \\ Centro de investigaci\'on CITIC, Databases Lab.}
  \streetaddress{Campus de Elvi\~na s/n}
  \city{A Coru\~na}
  \state{Spain}
  \postcode{15071}
}

\author{Guillermo de Bernardo}
\email{gdebernardo@udc.es}
\affiliation{%
  \institution{Universidade da Coru\~na, \\ Centro de investigaci\'on CITIC, Databases Lab.}
   \city{A Coru\~na}
  \state{Spain}
}

\author{Gonzalo Navarro}
\email{gnavarro@dcc.uchile.cl}
\affiliation{%
  \institution{IMFD, DCC, University of Chile}
  \city{Santiago}
  \state{Chile}
}

%
\renewcommand{\shortauthors}{Brisaboa et al.}

%
\begin{abstract}
We introduce a new family of compressed data structures 
to efficiently store and query large string dictionaries in main
memory. Our main technique is a combination of hierarchical 
Front-coding with ideas from longest-common-prefix computation in 
suffix arrays. Our data structures 
yield relevant space-time tradeoffs in real-world dictionaries. We focus
on two domains where string dictionaries are extensively used and efficient
compression is required: URL collections, a key element in Web graphs and
applications such as Web mining; and collections of URIs and literals, 
the basic components of RDF datasets. Our experiments show that our data
structures achieve better compression than the state-of-the-art alternatives
while providing very competitive query times.

\end{abstract}

%
%


%

%
\maketitle

\section{Introduction}

A string dictionary is essentially a bidirectional mapping between strings and 
identifiers. Those identifiers are usually consecutive integer numbers that
can be interpreted as the position of the string in the dictionary.
By using string dictionaries, applications no longer need to store
multiple references to large collections of strings. Replacing those strings, 
which can be long and have different lengths, with simple integer values
simplifies the management of this kind of data.

Many applications need to make use of
large string collections. The most immediate ones are text
collections and full-text indexes, but several other applications,
not specifically related to text processing, still require an
efficient representation of string collections. Some relevant examples include
those handling Web graphs, ontologies and RDF datasets, or biological sequences.
Web graphs, for example, store a graph representing hyperlinks between Web
pages, so the node identifiers are URLs. Most representations transform those
strings into numeric identifiers (ids), and then
store a graph referring to those ids. Compact Web graph representations can
store the graphs within just a few bits per edge~\cite{webgraph, ktree}.
Since the average node arities are typically 15--30, storing the URL of the
node becomes in practice a large fraction of the overall space.
In RDF datasets, information is stored as a labeled graph where nodes are either
blank, URIs, or literal values; labels are also URIs. The usual approach to
store RDF data is also to use string dictionaries to obtain numeric identifiers
for each element, in order to save space and speed up queries~\cite{rdfx}.
The classical technique of storing a string dictionary is extended in
some proposals by keeping separate dictionaries for URIs and literal
values~\cite{rdfdict}.

In this paper we consider the problem of efficiently storing large static
string dictionaries in compressed space in main memory, providing efficient
support for two basic operations: \lookup[s] receives a string and returns the
string identifier, an integer value representing its position in the dictionary;
\access[i] receives a string identifier and returns the string in the dictionary
corresponding to that identifier.

We focus on two types of dictionaries that are widely used in practical
applications: URL dictionaries used in Web graphs, which are of special interest
for many Web analysis and retrieval tasks; and URIs and literals
dictionaries for RDF collections, which are a key component of the Web of Data
and the Linked Data initiative and have experienced a sharp growth in recent
years. 

Our techniques achieve compression by exploiting
repetitiveness among the strings, so they are especially well suited to URL and
URI datasets where individual strings are relatively long and very similar to
other strings close to them in lexicographical order. In particular, we build
on Front-coding, which exploits long common prefixes between consecutive 
strings, and design a hierarchical version that enables binary searches without
using any sampling. We enhance this binary search with techniques inherited
from suffix array construction algorithms, which boost the computation of 
longest common prefixes along lexicographic ranges of strings. These main
ideas are then composed with other compression techniques.

Experimental results on real-world datasets show that our data structures
achieve better compression than the state-of-the-art alternatives, and we are
much faster than the few alternatives that can reach similar compression. Even
if faster solutions exist, our techniques are still competitive in query times
and significantly smaller than them.

The remaining of this paper is organized as follows: in
Section~\ref{sec:relatedWork} we introduce some concepts and refer to previous
work in string dictionary compression. Section~\ref{sec:proposal} presents our
proposal, describing the structure and query algorithms and explaining the main
variants implemented. Section~\ref{sec:experiments} contains the experimental
evaluation of our structures. Finally, Section~\ref{sec:conclusions} summarizes
the results and shows some lines for future work.

\section{Related work}
\label{sec:relatedWork}

\subsection{Previous concepts and basic compression techniques}

In this section we introduce some preliminary concepts, presenting existing data
structures and compression techniques that are used in the paper.

\subsubsection{Bit sequences}

A bit sequence or bitmap is a sequence $B[1,n]$ of $n$
bits. Bit sequences are widely used in many compact data structures. Usually,
bit sequences provide the following three basic operations:  $access(B,i)$
obtains the value of the bit at position $i$, $rank_v(B,i)$ counts the number of bits
set to $v$ up to position $i$, and $select_v(B,j)$ obtains the position in $B$
of the $j$-th bit set to $v$. All the operations can be answered in constant
time using $n+o(n)$ bits~\cite[Ch. 4]{cds}. Additionally, compressed bit
sequence representations have been proposed to further reduce the space
requirements~\cite{rrr}.
In this paper we use an implementation of the SDArray compressed
bitmap~\cite{sdarray} provided by the Compact Data Structures
Library~\texttt{libcds}\footnote{https://github.com/fclaude/libcds}. This
solution can achieve compression when the sequence is sparse and still
supports $select$ queries in constant time.

\subsubsection{Integer compression techniques}
In this paper we use Variable-byte (Vbyte) encoding~\cite{vbyte}, a simple
integer compression technique that essentially splits an integer in 7-bit
chunks, and stores them in consecutive bytes, using the most significant bit of
each byte to mark whether the number has more chunks or not. It is simple to
implement and fast to decode.

A technique of special relevance is Directly Addressable Codes
(DACs)~\cite{dacs}.
This technique aims at storing a sequence of integers in compressed space while
providing direct access to any position. Given the Vbyte encoding of the
integers, DACs store the first chunk of each integer consecutively, and use a
bitmap $B_1$ to mark the entries with a second chunk. The process is repeated
with the second chunks and its corresponding bitmap $B_2$, and so on. DACs
support decompressing entries accessing the first chunk directly and using
$rank_1$ operations on the $B_i$s to locate the corresponding position of the
next chunk.

DACs can work with Vbyte encoding but they are actually a general
chunk-reordering technique.
In this paper we make use of a variant that is designed to store a collection of
variable-length integer sequences, instead of a sequence of integers. In this
variant, that we call DAC-VLS, integers are not divided in chunks; instead, the
first integer in each sequence is stored in the first level, and a bitmap is used to mark whether the current
sequence has more elements. This technique does not reduce the space of the
original integers, but provides direct access to any sequence in the collection.

\subsubsection{String compression: Front-coding and Re-Pair}

Front-coding is a folklore compression technique
that is used as a building block in many well-known compression algorithms. Front-coding
compresses a string $s$ relative to another $s_0$ by computing their
longest common prefix (lcp) and removing the first $lcp$ characters from
the encoded string. Hence, Front-coding represents $s$ as a tuple containing the
$lcp$ and the substring after it $\langle lcp, s[lcp..len(s)] \rangle$.
Despite its simplicity, it is a very useful technique for many applications,
providing a simple way to compress collections of similar strings. URLs, for
instance, tend to have relatively long common prefixes, so Front-coding
compression is very effective on them, even if the string portions remaining
after Front-coding, or string tails, are still relatively long.

Re-Pair~\cite{repair} is a grammar compression technique that achieves good
compression in practice for different kinds of texts. Given a text $T$, Re-Pair
finds the most repeated pair of consecutive symbols $ab$ and replaces each occurrence
of $ab$ by a new symbol $R$, adding to the grammar a new rule $ R \rightarrow
ab$. The process is repeated until no repeated pairs appear in the text. The
output of Re-Pair is a list of $r$ rules and the resulting reduced text $T^C$,
represented as a sequence of integers in the range $(1,\sigma+r)$, where
$\sigma$ is the number of different symbols in the original text.

\subsection{String dictionary compression}
\label{subsec:sdc}

Simple techniques for storing collections of strings have been used in many
applications. Hash tables and tries~\cite{knuth1998} are just some examples of
classical representations that can be used in main memory for small
dictionaries.

As the dictionary size increases, those classical data structures no longer fit
in main memory, so a compressed representation has to be used or the dictionary
must be stored in secondary memory. A simple approach to
reduce space is to compress individual strings using general or domain-specific
compression techniques, before adding them to the dictionary structure.
Modern techniques for dictionary compression are based on specific compact data
structures usually combined with custom compression techniques applied to the
strings. Several theoretical solutions have been proposed for
static dictionaries~\cite{bille}, and solutions also exist for the dynamic dictionary
problem~\cite{dpct, kanda1, kanda2}. In this section we will focus on practical
solutions for a static dictionary, outlining the most relevant
existing implementations.

Martinez-Prieto et al.~\cite{Martinez-Prieto:2016} have proposed a collection of
compressed string dictionary representations that provide a choice for
different space/time tradeoffs.
In their survey, they show advantages against proposals based on compressed
tries and similar compression techniques. Their representations are based on
well-known compression techniques that are combined to build space-efficient
versions of data structures like tries and hash tables.
The most relevant proposal in this survey is a collection of differentially
encoded dictionaries.
The authors sort the strings and split them into fixed-size buckets. Then, they
store the first string of each bucket, or bucket header, in full, and the
remaining strings of the bucket are compressed relative to the previous one
using Front-coding. To answer \lookup queries, a binary search in the
bucket headers is used to locate the bucket containing the string, and a
sequential search in the bucket is performed; \access queries just traverse
sequentially the bucket containing the query identifier.
The authors propose several variants of this idea in the
original paper that combine the previous idea with additional
compression techniques like Huffman~\cite{huffman}, Hu-Tucker~\cite{hutucker}
or Re-Pair applied to the strings in each bucket or to the bucket headers to reduce the overall space usage.

In the previous work several other alternatives are proposed that share
similarities with our proposal. Binary-searchable Re-Pair (RPDAC) compresses the
strings with Re-Pair and uses DAC-VLS to provide direct access to each one,
supporting \lookup queries through binary search.
An improvement on the same idea uses a hash table to provide direct
access to the location of a string, instead of resorting to binary search, improving \lookup queries
significantly at the cost of additional space.

Grossi and Ottaviano propose a structure based on path
decomposed tries (PDT)~\cite{pdtries}. The authors create a path decomposition of the trie representing the dictionary
strings, and build a compact representation of the tree generated by the path
decomposition. They explore different techniques for the
representation of the trie (lexicographical and centroid-based
path decomposition). They also propose compressed variants in which the
path labels are compressed using Re-Pair.
Their solution has shown good results in different kinds of string dictionaries. Their compressed tries are
competitive in space with previous techniques, but more importantly provide fast and very consistent query times.

Arz and Fischer~\cite{lzstringdict} have recently proposed a solution based on
Lempel-Ziv-78 (LZ-78) compression on top of PDT. This technique has been shown
to slightly improve the compression of PDT in some datasets, but improvement is
small in most cases and the LZ-78-compressed structures have much higher query
times, especially in \lookup queries.

\section{Our proposal}
\label{sec:proposal}

\subsection{Data structure and algorithms}

We propose a family of compression techniques for string collections that aim at
providing good compression with efficient query times. Our techniques follow
some of the ideas of differential compression described in
Section~\ref{subsec:sdc} and aim at improving their weak points.

To build our representation, the strings are sorted in lexicographic order.
This order is frequently used in most string dictionary representations, so
that entries that are close to each other should also be similar to each other.
For convenience, we also add two marker strings at the beginning and at the end
of the collection: the former is the empty string, and the latter is a
single-character string lexicographically larger than any string in the original collection.

Our goal is to use Front-coding to reduce the common prefix of
common entries. However, instead of compressing each string relative to the
previous one, we use a different scheme for comparisons that constitutes the
basis of our proposal. Our technique is based on a binary decomposition of the
list of strings, following similar ideas to the binary search algorithms over
suffix arrays proposed by Manber and Myers~\cite{suffixArray}.

Assume we have a collection $C$ of $n$ strings, including our initial and
last string, and let $C[\psubi]$ be the string at position $\psubi$ in the
collection.
Our structure is built as follows:
\begin{itemize}
  \item We initialize two markers $\psubl=0$ and $\psubr=n-1$, set to the limits of
  the collection.
  \item We select the middle point $\psubm = (\psubl+\psubr)/2$ and compute
  $\llcp[m]=\lcp(C[\psubm], C[\psubl])$ and $\rlcp[m]=\lcp(C[\psubm],
  C[\psubr])$, the longest common prefixes between the string at position $\psubm$
  and the strings at both limits of the interval.
  \item Let $\maxlcp$ be the maximum between $\llcp[\psubm]$ and $\rlcp[\psubm]$.
  $C[\psubm]$ is compressed using Front-coding, by removing the $\maxlcp$
  initial bytes. In practice, Front-coding is applied relative to the most similar of
  the entries at each limit of the interval. We will refer to these as the
  ``parents'' of a given entry.
  \item We recurse on both halves of the collection ($[0,\psubm]$ and $\psubm,n-1]$),
  repeating the previous steps to compare the middle element with the limits of
  the interval and apply Front-coding accordingly.
\end{itemize}

After this procedure, our conceptual representation consists of two
integer sequences $\llcp$ and $\rlcp$, and the remaining of each string after Front-coding is applied to
them. Let us call this $S[n]$. In practice we use different
techniques to store the strings, but for simplicity we will write
$S[i]$ to refer to the string stored at position $i$.

\begin{figure*}[h]
  \centering
  \includegraphics[width=\linewidth]{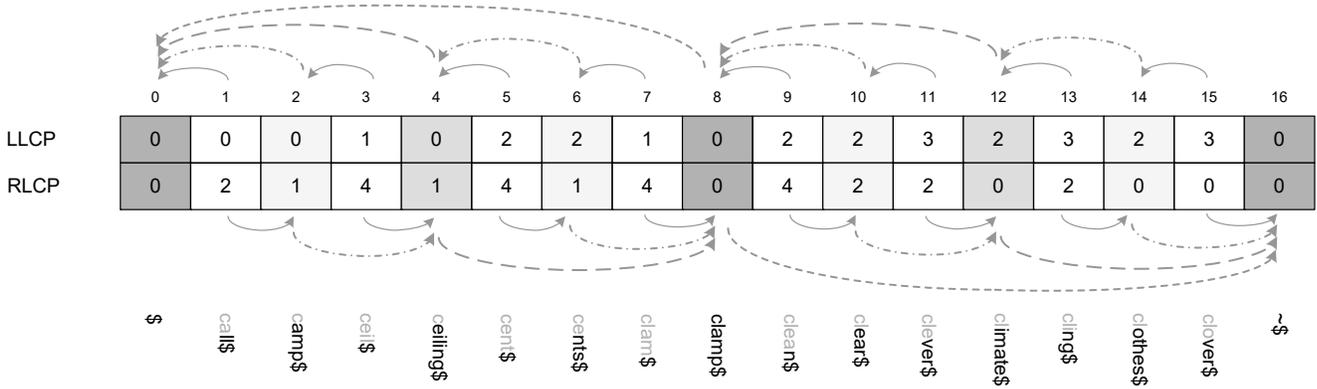}
  \caption{Conceptual dictionary structure. The original strings are displayed below, but the grayed-out prefixes are not stored.}
  \label{fig:structure}
  \Description{Conceptual dictionary structure. The original strings are displayed below, but the grayed-out prefixes are not stored.}
\end{figure*}

Figure~\ref{fig:structure} shows an example of our dictionary structure for a
small set of strings. We use \texttt{\$} to denote a string terminator. Our
marker strings are denoted as \texttt{\$} and \texttt{\textasciitilde\$} respectively.
The original strings at each position are displayed below the arrays, with the
prefix that would be removed after Front-coding compression grayed out.
Arrows identify the position of the left and right ``parent'' of each entry. For
instance, $C[8]$ is compared with positions 0 and 16 (our marker strings), and it is stored in full. $C[12]$ (\texttt{climate \$}) is compared with $C[8]$
($\llcp=2$) and $C[16]$ ($\rlcp[8]=0$), and after Front-coding is applied it becomes \texttt{imate\$}, removing the longest common prefix. Note that the
marker strings we use will never share a common prefix with any string in
the collection, so both marker strings and the string in the middle position
will always have \llcp and \rlcp values of 0 and will be stored in full. The
final representation needs to store the \llcp and \rlcp arrays and the collection of string tails.

Our construction technique is expected to yield worst compression results than
the usual Front-coding approach that would be applied sequentially to the
collection of strings. We will describe later our implementation strategies to
improve the space utilization. However, as we will see next,
our binary decomposition allows us to provide an efficient method to answer queries without resorting to
sampling or partitioning of the collection. Therefore, we avoid the need for bucket headers that 
arises in some of the solutions described in Section~\ref{sec:relatedWork}.

Next we outline the algorithms for \lookup and \access operations. In both cases
we perform a binary-search-like traversal of the collection.

\subsubsection{Lookup operation}

To obtain the identifier of a string in the dictionary (\emph{lookup}), a trivial
algorithm would involve a binary search, checking the midpoint at each step and comparing the
resulting string with the target. However, our scheme is able to improve the
performance of \lookup operations by avoiding some string comparisons.

The pseudo-code used for \lookup searches is described in Algorithm~\ref{alg:stringToId}.
Let $s_q$ be the search string.
The values $\psubl$ and $\psubr$ are the limits of our interval, initially $\psubl =
0$ and $\psubr = n-1$. The variables $\xll$ and $\xrr$ store the longest common
prefix of the left- and right-hand strings in the dictionary with the search
string and are initially set to 0.
Hence, a \lookup[\sq] is translated into $doLookup(\sq, 0, n-1, 0, 0)$.

At any step of search, we first compare $\xll$ and $\xrr$. We will focus on the
case $\xll >= \xrr$ (i.e., the string at $\psubl$ is more similar to $\sq$ than the string at $\psubr$) covered in lines 3-18 of the algorithm~\footnote{In
practice, when \xll = \xrr we have to check the values of \llcp and \rlcp to
choose the branch for traversal. Algorithm~\ref{alg:stringToId} shows the
actual comparison.}, since the other case is symmetric.
We obtain the midpoint $\psubm$ and the value of $\llcp[\psubm]$ and then compare it with
$\xll$:
\begin{itemize}
  \item If $\llcp[\psubm] > \xll$, entry $\psubm$ has a longer prefix in
  common with $\psubl$ than $\psubl$ with $\sq$. Hence, the result cannot be to
  the left of $\psubm$. We recurse on the right half of the range ($[\psubm, \psubr]$) without
  comparing strings.
  \item If $\llcp[\psubm] < \xll$, we are on the symmetric case: entry $\psubl$
  is more similar to the pattern than to entry $\psubm$. We recurse on the
  left half of the interval, and we set the new lower bound $\xrr =
  \llcp[\psubm]$, since our current string must have $\llcp[\psubm]$ characters
  in common with the search string.
  \item If $\llcp[\psubm] = \xll$, we need to compare entry $\psubm$ with $\sq$.
  Our comparison method in Algorithm~\ref{alg:stringToId} gives us the two
  relevant pieces of information: the comparison value $cmp$, and the offset
  $\off$ of the last equal character. If both strings are equal, we return
  immediately. Otherwise, we recurse on the appropriate half, setting the value of \xll
  or \xrr to $\off$.
\end{itemize}

\begin{algorithm}
\caption{Algorithm for \lookup}
\label{alg:stringToId}
\begin{algorithmic}[5]

\Function{doLookup}{$\sq$, $\psubl$, $\psubr$, $\xll$, $\xrr$}
	\State $\psubm \gets (\psubl + \psubr) / 2$

	\If { $\xll > \xrr$ \textbf{or} $((\xll = \xrr) \And \llcp[\psubm] >=
	\rlcp[\psubm])$} \State $lval \gets \llcp[\psubm]$
		\If { $lval > \xll $ }
			\State \Return \Call{doLookup}{$\sq$, $\psubm$, $\psubr$, $\xll$, $\xrr$}
		\ElsIf { $lval < \ell $ }
			\State \Return \Call{doLookup}{$\sq$, $\psubl$, $\psubm$, $\xll$, $lval$}
		\Else
			\State $(cmp, \off) \gets compare(s + \xll, S[\psubm])$
			\If {$cmp > 0$}
				\State \Return \Call{doLookup}{$\sq$, $\psubm$, $\psubr$, $\off$, $\xrr$}
			\ElsIf {$cmp < 0$}
				\State \Return \Call{doLookup}{$\sq$, $\psubl$, $\psubm$, $\xll$, $\off$}
			\Else
				\State \Return $\psubm$
			\EndIf
		\EndIf
	\Else
		\State $rval \gets \rlcp[\psubm]$
		\If { $rval > \xrr $ }
			\State \Return \Call{doLookup}{$\sq$, $\psubl$, $\psubm$, $\xll$, $\xrr$}
		\ElsIf { $lval < \xll $ }
			\State \Return \Call{doLookup}{$\sq$, $\psubm$, $\psubr$, $rval$, $\xrr$}
		\Else
			\State $(cmp, \off) \gets compare(s + \xrr, S[\psubm])$
			\If {$cmp > 0$}
				\State \Return \Call{doLookup}{$\sq$, $\psubm$, $\psubr$, $\off$, $\xrr$}
			\ElsIf {$cmp < 0$}
				\State \Return \Call{doLookup}{$\sq$, $\psubl$, $\psubm$, $\xll$, $\off$}
			\Else
				\State \Return $m$
			\EndIf
		\EndIf
	\EndIf
\EndFunction
\end{algorithmic}
\end{algorithm}

Following the example in Figure~\ref{fig:structure}, assume we are searching
for string \texttt{clam}. First we compare with \texttt{clamp} (due to our
markers, in the first iteration a comparison is always performed). The
query string is smaller than $S[8]$, and they share the first 4 characters.
Therefore, we recurse on the left half $[0,8]$, setting $\xrr = 4$. In the next
step ($\psubm=4$), $\xrr > \xll$ and $\rlcp[4] = 1 < \xrr$, so we do not need to
compare strings: we just recurse on the right-side interval $[4,8]$, and
we set $\xll=1$, since $\rlcp[4]=1$, meaning that it shares also a prefix of
length 1 with \sq. In the next step ($\psubm = 6$), again $\xrr > \xll$, and
$\rlcp[6] = 1 < \xrr$, so we recurse on the interval $[6,8]$.
At the last step, $\rlcp[7] = 4 = \xrr$, so we compare strings to find that both
strings are equal.

\subsubsection{Access operation}

The second main operation, \access[i], is the opposite of the previous
one, retrieving the string for a given identifier. It follows a bottom-up
approach, starting at the position $\psubi$ and traversing up to the parent
position until we have recovered the full string. The procedure is described in Algorithm~\ref{alg:idToString}. The string is decoded from the
end, prepending new characters at each new step until we reach the beginning of
the string. Given an identifier $i$, we read $\llcp[\psubi]$ and $\rlcp[\psubi]$
and compute their maximum as $\off$. Then, we can extract all the characters from
$S[\psubi]$, that will correspond to the result string from position $\off$
onwards. Since we have already decoded the result from position $\off$, in the
next iterations we set a $limit$ to mark that we only need to extract characters
up to that position.

After extracting the required characters, we move to the appropriate
parent\footnote{In practice, the parent positions are not computed bottom-up in
our implementations. Instead, the list of search positions is obtained in a top-bottom fashion before the \access algorithm starts. These details are
omitted for simplicity in Algorithm~\ref{alg:idToString}.}, the one
corresponding to the maximum lcp, and repeat the procedure. Whenever $\off \leq
limit$, we prepend the first $limit - \off$ characters of the current $S[i]$ to
the result. When we reach $\off = 0$ the result has been decoded and
the procedure ends.

Note that in the worst case we may have to traverse up
until we reach one of the positions that are always stored in full: $0$, $(n-1)/2$ or $n-1$, hence running $\log(n)$ string comparisons.
However, in many instances we can reach $\off=0$ earlier in the traversal.
Additionally, in iterations where $\off > limit$ comparisons are skipped, therefore we
do not even need to access the text. This will be relevant in some
implementation variants that apply compression to the string tails, since in
those solutions string comparisons are relatively expensive.

\begin{algorithm}
\caption{Algorithm for \access}
\label{alg:idToString}
\begin{algorithmic}[5]

\Function{access}{$\psubi$}
	\State $s \gets ''$
	\State $limit \gets max(\llcp[\psubi], \rlcp[\psubi])+len(S[\psubi])$
	\State $\off \gets \infty$
	\While {limit > 0}
		\If {$\llcp[\psubi] \ge \rlcp[\psubi]$}
			\State $(\off, n) \gets (\llcp[\psubi], left(\psubi))$
		\Else
			\State $(\off, n) \gets (\rlcp[\psubi], right(\psubi))$
		\EndIf
		\If {$\off \le limit$}
			\State $s[\off .. limit] \gets S[\psubi][0 .. limit - \off]$
			\State $limit \gets \off - 1$
		\EndIf
		\State $\psubi \gets n$
	\EndWhile
	\State \Return $s$
\EndFunction
\end{algorithmic}
\end{algorithm}

Following again the example in Figure~\ref{fig:structure}, assume we want to
obtain the string for identifier 9 (\texttt{clean}). At the first iteration, the
maximum common prefix is $\rlcp[9]=4$.
This means that $S[4]$ is stored from position 4, so we can recover the
characters from position 4 until the end of string (\texttt{\_\_\_\_n}). We
set $limit = 4$ for future iterations, and since the \rlcp value was higher we
move to the right-side parent, i.e. to position 10.
Now $\llcp[10]=\rlcp[10]=2$, so $\off = 2$ and we can extract the first two
characters of $S[10]$ to fill positions 2-3 of the result string, getting
\texttt{\_\_ean}. In this step we could move to either side, assume we
simply move left by convention. We reach position 8, and we get
$\llcp[8]=\rlcp[8]=0$, so we copy the first two characters of $S[8]$ to fill
the remaining positions of our result and then return.

\subsection{Implementation variants}

Our conceptual representation stores two integer sequences $\llcp$ and $\rlcp$
and a set of string tails $S$. Several alternatives exist for the
representation of both structures, hence originating a family of structures that provide a space/time
tradeoff. In this section we introduce implementation details for the different
variants of our proposal:

\mmpfc is the simplest proposal. In this approach we store $\llcp$ and $\rlcp$
as sequences of fixed-length integers. This solution is simple and efficient, but in datasets
  where the maximum lcp value is high it is space-inefficient. The string tails
  $S$ are concatenated in a single sequence $Str$. A bitmap $B$ is added to
  indicate the position in $Str$ where each string begins marking with 1 those
  positions and setting the remaining positions to 0. We store the bit array using an SDArray
  compressed bitmap representation, to provide \emph{select} support. In this
  representation, $S[i]$ is obtained by selecting the position of the $i$-th 1
  in $B$, and extracting $Str[select_1(B,i) ..  select_1(B,i+1)-1]$.

\mmrpfc differs from the previous one on the
  representation of the strings. All the string tails are again concatenated in
  a single sequence $Str$, including the end-of-string markers, or string
  terminators.
  After this, a variant of Re-Pair compression is applied to the sequence, generating
  a grammar-compressed sequence where symbols never overlap two dictionary
  strings. This transforms the original byte string into a grammar and a
  sequence of integers. The sequence of integers is encoded using Vbyte. We
  also use a bitmap $B$ that marks with 1 the first byte of each dictionary
  string.
  $S[i]$ can be obtained by extracting the sequence in the same way as before,
  and then decoding the corresponding Re-Pair sequence.

\mmrpfcdac is similar to \mmrpfc but it uses DACs to store the sequences
$\llcp$ and $\rlcp$. This is expected to achieve much better space in
many real-world collections, and especially in collections with long
strings where the maximum lcp is much higher than the average.

\mmrpfcdacvls is again similar to \mmrpfc but uses the variant of DACs designed
for variable-length integer sequences (DAC-VLS) to store the Re-Pair-compressed
strings (i.e. the sequence of integers generated by Re-Pair), instead of compressing
individual integers with Vbyte. Since the DAC-VLS structure
provides direct access to any string, the bitmap $B$ is not necessary, and a string $S[i]$ is
just decoded by extracting symbols from the DAC-VLS structure and decompressing
them using the Re-Pair grammar. Note that this combination of Re-Pair and
DAC-VLS is the same underlying idea of RPDAC, described in
Section~\ref{sec:relatedWork}.

\mmrpfcdacvlsb combines the two previous ones: $\llcp$ and $\rlcp$ are
stored using DACs, and $Str$ stored using DAC-VLS.

\subsubsection{End-of-string symbols}

All our implementations use a bitmap $B$ or a DAC-VLS structure to provide
direct access to any string tail, so, unlike alternatives based on sequential
search, our representation does not need to physically store end of string
markers. The string  terminators are used as markers when applying Re-Pair
compression, so that no Re-Pair symbol overlaps two dictionary strings.
However, after compression, we can remove these string terminators to save a
byte per string in $Str$. Nevertheless, we still tested, as well, the version
with string terminators since having them we can decode until we reach the
terminator instead of performing a second $select_1$
operation on $B$. Even though $select$ operations are constant-time, they are
relatively costly and avoiding them we can speed up string decoding.

Notice that, when a string is compressed relative to a larger string in
lexicographical order, a zero-length tail may appear (see for example the
string at position 5 in Figure~\ref{fig:structure}). We handle these empty strings as a special case,
storing them as an end-of-string symbol even if our implementation would remove
these symbols in any other case. This is necessary for \emph{select} operations
in $B$ to work, so that each $S[i]$ is associated with a different offset in
$Str$; the DAC-VLS structure also requires this adjustment since it is not
designed to support zero-length sequences. Note also that the DAC-VLS
implementation, due to its construction, would not benefit from extra end-of-string
symbols, so for those implementations we only use variants with no string
terminators.

\subsubsection{Single-lcp implementations}

Our main proposal stores two integer sequences, $\llcp$ and $\rlcp$, to optimize
\lookup operations. Similar algorithms can be designed to work with only one
array, saving half the space of these arrays at the cost of worst Front-coding
compression.

Single-lcp implementations of any of our proposals can also be built in order to
reduce the space utilization. The same idea of the general construction applies
to these variants, but now we always compare with the left parent
(llcp-only variants) or with the right parent (rlcp-only variants). Compression
of the strings is expected to be worse since we are no longer using the
maximum lcp, but these variants can still achieve better overall compression by
removing one of the integer sequences.

Regarding query algorithms, \lookup operations can still save some
string comparisons using a similar algorithm to the one we proposed:
essentially, llcp-only variants use lines 4-18 of the original algorithm, and rlcp-only variants lines
20-34. On \access operations, the algorithm is also essentially the same,
but we always move to the left (right) parent. When the
lcp arrays are compressed, removing one access to them will have
a positive effect on performance, since a single-lcp implementation only needs one DAC access per
step.

\section{Experimental evaluation}
\label{sec:experiments}

In this section we test the performance of our proposal in comparison with
several alternatives in the state of the art. We perform tests with real-world
datasets, focusing on two main application domains: representation of URLs,
obtained from Web graph crawls, and representation of URIs and literal values extracted from
RDF datasets. First we show an empirical evaluation of the implementation
variants described in Section~\ref{sec:proposal}, in order to display their
strengths. Then, we perform an experimental evaluation of our best
implementation variants, comparing them with existing solutions
for string dictionaries. Our comparison focuses on
compression capabilities and query performance, and shows that our solutions
obtain a better trade-off than state-of-the-art alternatives.

\subsection{Experimental setup}

We use in our tests a collection of datasets including URLs from real Web
graphs and also URIs and literal values from an RDF dataset.
Table~\ref{tab:datasets} shows a summary of the datasets used. For each one,
we display its size in plain, the number of strings it stores,
the average string length and the alphabet size. Note that the average length displayed is
computed as total size divided by number of strings, so it includes an extra
character per string corresponding to the string terminator in the input.

\begin{table}
  \caption{Description of the datasets}
  \label{tab:datasets}
  \begin{tabular}{lrrrr}
    \toprule
    Dataset		&	Size(MB)	&	\#strings	&	Avg. length & $\sigma$ \\
    \midrule
   	\duk		&	1372.06		&	18,520,486	&	77.68 	&  101 \\
   	\darabic	&	1774.42		&	22,744,080	&	81.81	&  100 \\
   	\duris		&	1553.46		&	30,137,450	&	54.05	&  116    \\
   	\dliterals	&	2048.00		&	331,253,572	&	7.48	&  96    \\
  \bottomrule
\end{tabular}
\end{table}

\emph{\duk} and \emph{\darabic} are datasets containing URLs of two different
Web graph crawls. \duk\footnote{http://law..dsi.unimi.it/webdata/uk-2002} has been
obtained from a 2002 crawl of \texttt{.uk} domains, whereas
\darabic\footnote{http://law.di.unimi.it/webdata/arabic-2005/} is a 2005 crawl
that includes pages from countries whose content is potentially written in Arabic.
Both datasets have been obtained from the Webgraph framework~\cite{webgraph}. The
\duk dataset has been used in previous work as a baseline for URL
compression~\cite{Martinez-Prieto:2016, pdtries, lzstringdict}. The \darabic
dataset is included for better confirmation of the performance
of each solution in different Web graphs. Both datasets are similar in
number of strings and average string length.

\emph{\duris} contains all the different URIs in the English version of the
DBpedia RDF dataset, in its 3.5.1 version\footnote{http://downloads.dbpedia.org/3.5.1/all\_languages.tar}.

\emph{\dliterals} is a subset of the literals existing in the same DBpedia 3.5.1
dataset. Our input was generated from the original data by extracting
all the literal values of the collection and obtaining the raw value from the
RDF literal. To do this we remove language tags and type information, as
well as the enclosing quotes of the original string. For instance, the RDF
literal
\texttt{\small \textquotedbl{}100 AD\textquotedbl{}@en}
becomes \texttt{\small 100 AD} after removing the language tag, whereas the
numeric value \texttt{\small
\textquotedbl{}57805\textquotedbl{}\^{}\^{}<http://www.w3.org/2001/XMLSchema\#int>}
is converted to \texttt{\small 57805}. We sorted the values lexicographically,
discarding duplicates and taking the entries in the first 2 GB. We limit the input size to 2 GB since it is the maximum supported by most of the state-of-the art alternatives that will be used for comparison.
Notice that using raw literal values the strings are significantly shorter, but
keeping the full strings would have little effect on our techniques: since only
one language tag and a small number of different types are used, Re-Pair
compression would be able to represent the extra characters at small
cost. Our choice of raw values aims at highlighting the fundamental differences
between \dliterals and the other datasets used, as \dliterals has much
shorter strings on average, and much more different from each other.

The space shown for each structure is computed precisely from the size of the
corresponding components. To measure query times, we build a set of 10,000
queries for each dataset by selecting random positions from the collection. The
same positions are used for \access and for the corresponding \lookup queries.
Query times are measured as the average over 100 iterations of the query set.

We implemented our proposals in C++\footnote{Our code is publicly available at
https://gitlab.lbd.org.es/gdebernardo/improved-csd}. We use an
implementation of compressed bitmaps and Re-Pair based on the \texttt{libcds} library, the same
used by Martinez-Prieto et al.~\cite{Martinez-Prieto:2016}. All our
implementations are compiled with g++ 4.8 with -O9 optimizations.

We compare our results with the following techniques:

\begin{itemize}
  \item \pfc, \rpfc and \rphtfc are some of the differential encoding techniques
  based on Front-coding~\cite{Martinez-Prieto:2016} described
  in Section~\ref{sec:relatedWork}. \pfc is the plain solution, \rpfc uses
  Re-Pair to compress buckets. \rphtfc is similar to the previous one, but
  it also applies Hu-Tucker compression to the bucket headers. We include \pfc
  because it is the simplest solution, and \rpfc and \rphtfc because they
  achieved the best results among their Front-coding-based solutions. We used
  bucket sizes 4, 8, 16 and 32.
  \item \rpdac and \hashrp are the binary searchable Re-Pair techniques also
  introduced in Section~\ref{sec:relatedWork}. The first one uses binary
  search, and the second one adds a hash table to speed up queries. Both of them
  are used with the default configuration parameters.
  \item \pdt is the the centroid-based compressed
  implementation of path-decomposed
  tries variants~\cite{pdtries}, the best-performing alternative of this family.
\end{itemize}

All the alternatives are compiled with g++ with full optimizations enabled,
using the default settings as provided by the authors apart from the parameters
described above.

Note that we do not include a comparison with the
implementation of LZ-78-compressed tries also described in
Section~\ref{sec:relatedWork}, since their publicly-available code could not be
compiled. Nevertheless, previous results~\cite{Martinez-Prieto:2016,lzstringdict} suggest that their proposals are
dominated in most cases by \pdt, and when they slightly improve compression they
are much slower; they are also less efficient than \hashrp
and \rphtfc in most cases.

\subsection{Comparison of our variants}

Due to the relatively large number of variants proposed, we first outline some
of the general characteristics of our implementation variants to display their
relative strengths. After that, in the following sections we will only show experimental results corresponding to those of our techniques
that provide the best tradeoff.

\begin{figure}[h]
  \centering
  \includegraphics[width=\linewidth]{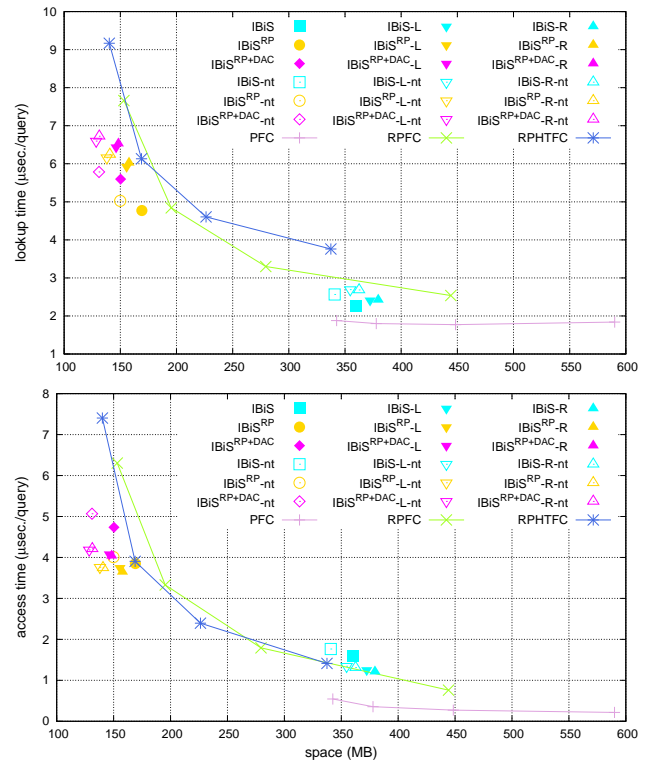}
  \caption{Comparison of our variants on dataset \duk}
  \label{fig:compours}
  \Description{Comparison of implementation variants on dataset \duk}
\end{figure}

Figure~\ref{fig:compours} shows the space/time tradeoff provided by some of our
proposals, considering both uncompressed (\mmpfc) and Re-Pair-compressed
strings (\mmrpfc, \mmrpfcdac). For each approach we show the space/time
tradeoff achieved in the dataset \duk for the basic implementation (two lcp
arrays) and both possible single-lcp implementations, labeled \texttt{-L} and
\texttt{-R} respectively. For each of those, we show results for the basic
techniques that keep string terminators and also for \vsinceros implementations
(labeled with \texttt{-nt}). The plot also shows a few of the
differential encoding techniques described in Section~\ref{sec:relatedWork},
since they share similarities with our approach:
\pfc is similar to \mmpfc, whereas the rest of our variants are similar to \rpfc or \rphtfc, improving compression through the use of
Re-Pair and other techniques.

As shown in Figure~\ref{fig:compours}, our plain implementations are not
very competitive with \pfc, since our techniques cannot compress each string
with Front-coding as efficiently as the sequential encoding. Plain approaches
will be omitted in the next sections, focusing on the more space-efficient
alternatives. Figure~\ref{fig:compours} also shows some trends among our
variants that are mostly the same in all the datasets used in our experiments:
\begin{itemize}
  \item Single-lcp implementations are, in general, a bit more
  space-efficient than double-lcp implementations in all the variants that use Re-Pair. A single-lcp
  variant may have significantly more characters in the string tails than a
  double-lcp variant. However, due to the efficiency of Re-Pair to compress the resulting strings, the actual increase in size of the
  compressed text is much smaller, and removing one of the lcp arrays
  easily compensates for this additional space. Plain single-lcp
  implementations, on the other hand, are much less efficient in space, since
  the extra bytes in the string tails are not compressed in any way.
  Regarding query times, single-lcp implementations are slower on \lookup queries,
  due to the potentially larger cost of searches, but faster on \access queries,
  thanks to the simpler bottom-up traversal that only needs to access a single
  lcp array. We will show experimental results for both single-lcp and
  double-lcp variants, since they can be useful in different scenarios depending
  on whether \lookup or \access queries are more relevant.
  \item \vllcp and \vrlcp implementations achieve
  almost identical query times, as expected. However, \vllcp achieves slightly better
  compression in all cases, and it is also simpler, since in \vllcp
  variants we always perform Front-coding compression respective to a
  lexicographically smaller string, so we are guaranteed to have non-empty
  string tails in every position. In view of these results, we will omit \vrlcp
  variants from the remaining test results, noting that in all our experiments
  they were consistently slightly larger than their \vllcp counterparts and
  query times are similar.
  \item \vsinceros implementations achieve much better compression in most
  variants and in all datasets. This is expected since after Re-Pair
  compression is applied to $Str$ the average length of a string tail is usually much shorter, so removing a
  byte per word yields a significant reduction in the overall space. As
  expected, \vsinceros variants are also slightly slower, both in \lookup and
  \access queries, but we consider the effect on compression much more relevant.
  In the remaining test results we will focus mostly on \vsinceros variants.
\end{itemize}

\subsection{Comparison with the state of the art}

Next we compare our implementations with the most significant
state-of-the-art alternatives to the best of our knowledge. Note that, as stated
earlier, we omit some of our implementation alternatives to provide clearer
plots, and focus our comparison on the best-performing techniques from
previous work.

\begin{figure}[ht]
  \centering
  \includegraphics[width=\linewidth]{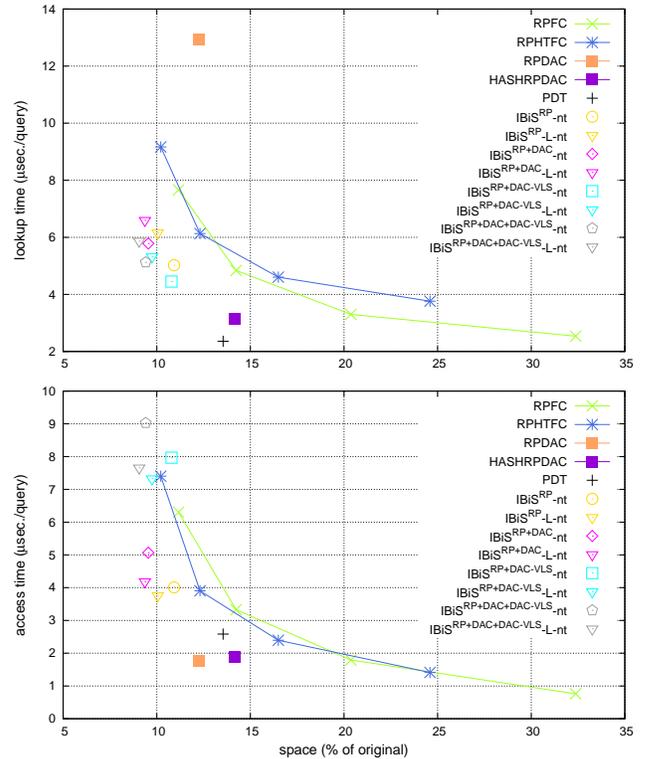}
  \caption{Space and query times on dataset \duk}
  \label{fig:compuk}
  \Description{Space and query times on dataset \duk}
\end{figure}

\begin{figure}[ht]
  \centering
  \includegraphics[width=\linewidth]{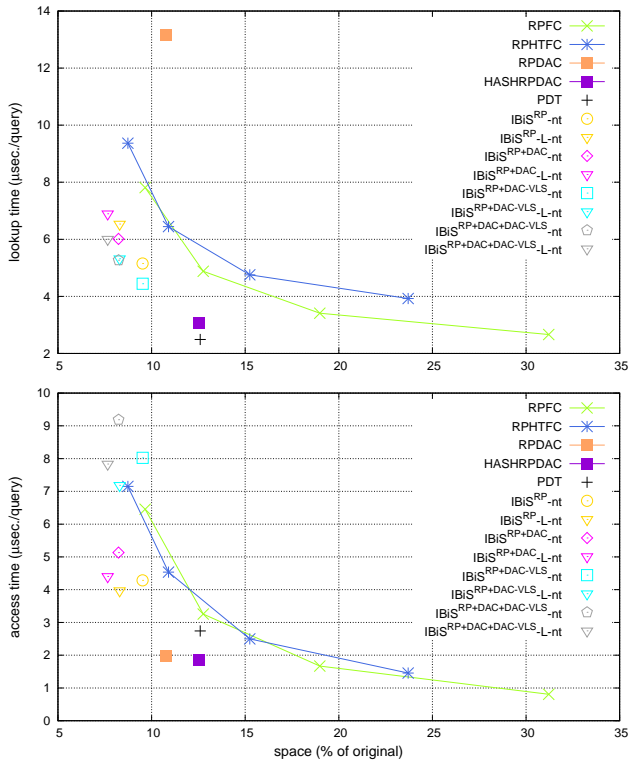}
  \caption{Space and query times on dataset \darabic}
  \label{fig:comparabic}
  \Description{Space and query times on dataset \darabic}
\end{figure}

Figures~\ref{fig:compuk} and~\ref{fig:comparabic} show the space/time tradeoff
on the Web graph datasets \duk and \darabic. Both datasets are similar and the
results obtained by the different techniques are also similar. Our proposals
achieve the best compression among all the tested implementations. The \vllcp variant of \mmrpfcdacvlsb obtains the best
overall space results, but the equivalent \mmrpfcdac is very close.
We improve the space/time tradeoff \rpfc and \rphtfc, since for smaller buckets they need much space to store bucket headers
and for larger buckets their sequential traversal of the bucket becomes much
slower.
Regarding query times, the most efficient techniques are \pdt and \hashrp; \rpdac is similar to \hashrp on \access queries, but much less competitive on
\lookup queries, since it requires a binary search and must decode an entry of
the DAC-VLS structure at each step.
Our variants are similar on \lookup queries, but the DAC-VLS solutions are
slower on \access queries. Note that our DAC-VLS solutions are much faster in
\lookup queries than \rpdac; both perform a binary search with
accesses to a DAC-VLS structure, but we encode shorter entries thanks to
Front-coding and we do not need to access the DAC-VLS at each step.
The query times of our best solutions are roughly two times slower than the
fastest solutions, but we are also significantly smaller than those, becoming the best alternative to optimize compression with competitive query times.

\begin{figure}[h]
  \centering
  \includegraphics[width=\linewidth]{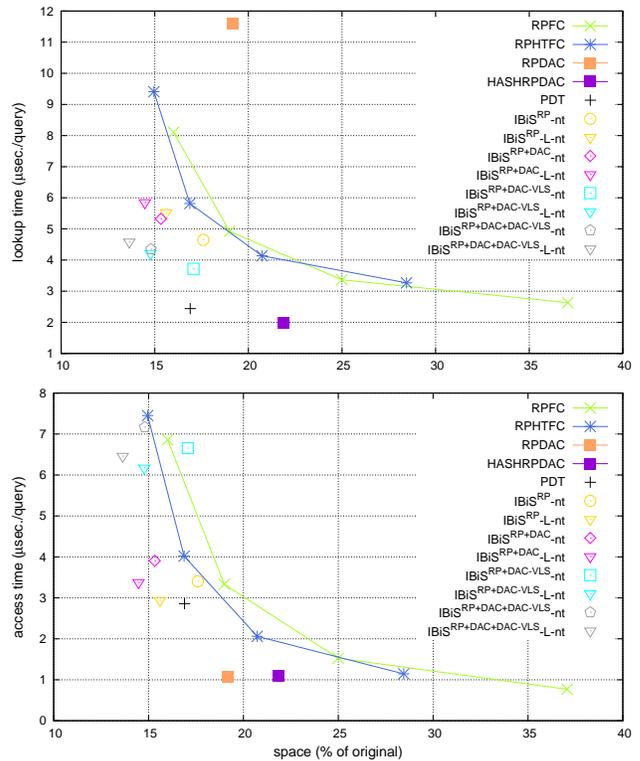}
  \caption{Space and query times on dataset \duris}
  \label{fig:compuris}
  \Description{Space and query times on dataset \duris}
\end{figure}

Figure~\ref{fig:compuris} shows the results for the \duris dataset. Overall
compression of all the tested representations is slightly worse when compared
with the URL collections, but our compressed representations achieve again the
best space results, around 15\% compression. Again, the best compression is
achieved by the \vllcp variants of \mmrpfcdacvlsb and \mmrpfcdac, and most of
our proposals improve the tradeoff provided by \rpfc and \rphtfc.
Our best variants are also significantly smaller than \hashrp, that achieves
the best query times. \pdt reaches compression close to ours, while
achieving better query times on \lookup queries. Nevertheless, on \access queries our best structures are still competitive with \pdt,
achieving similar query times in less space.
We consider this result on \access queries more relevant than the result on
\lookup queries since, in practice, the former are usually more relevant
than the latter, because they are more frequently used.
In an RDF engine, for instance, a SPARQL query just requires a few \lookup
operations to encode the URIs/literals used in the query into numeric
identifiers; then, after the query is executed, each result has to be translated
back into the corresponding URIs/literals, which means a potentially very large number of
\access operations to answer a single query.
Hence, even though good performance is required on both operations,
performance on \access queries may be more important in many applications.

\begin{figure}[h]
  \centering
  \includegraphics[width=\linewidth]{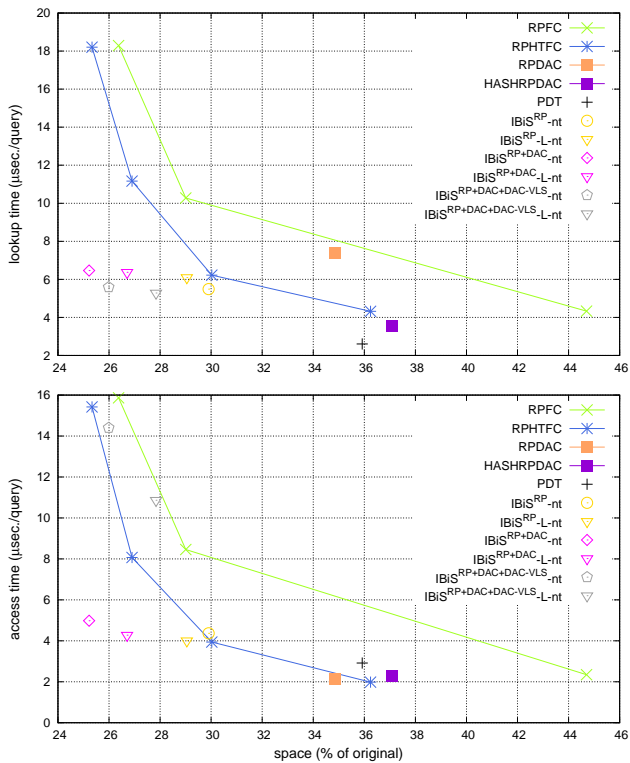}
  \caption{Space and query times on dataset \dliterals}
  \label{fig:compliterals}
  \Description{Space and query times on dataset \dliterals}
\end{figure}

Figure~\ref{fig:compliterals} shows the results obtained for the \dliterals
dataset. In this dataset, the different nature of the strings leads to
significantly different results: \pdt, \rpdac and \hashrp are much less
efficient to compress the collection. Also, among our variants, the DAC-VLS
techniques become much less efficient, since they are not well-suited to
handle this kind of collection, with very short average string length but a few
very long strings. Nevertheless, we still show in the plot the results for the
best performing DAC-VLS variants, namely the \mmrpfcdacvlsb approaches.
Note also that, in this dataset, \vllcp implementations are not as
efficient, and the smallest representation is the double-lcp \mmrpfcdac. In spite of all these differences, our best solutions (both double-lcp and single-lcp) are much smaller than \pdt and \hashrp, while obtaining query times competitive with them.
\rpfc and \rphtfc, for larger bucket sizes, can achieve compression similar to
us, but at the cost of much larger query times. Notice
that, due to the characteristics of this dataset, the overall compression of
all the solutions for this collection is much worse than in the previous ones,
but still \mmrpfcdac reaches 25\% compression whereas \pdt and \hashrp are above
35\%.

Taking into account the combined results from Figures~\ref{fig:compuris}
and~\ref{fig:compliterals}, our techniques clearly obtain the best compression
for both URIs and literal values, constituting a very efficient basis
for string dictionary compression of RDF data. Our query times are competitive
with those of existing data structures, especially on \access queries, and the
space-time tradeoff provided overcomes the tested alternatives.

\section{Conclusions and future work}
\label{sec:conclusions}

We have introduced a new family of compressed data structures for the
efficient in-memory representation of string dictionaries. Our solutions 
can be regarded as an enhanced binary search that combines a hierarchical 
variant of Front-coding with suffix-array-based techniques to speed up 
longest-common-prefix computations. Those ideas are then composed with
other techniques to derive a family of variants.

We perform a complete experimental evaluation of our proposals, comparing them
with the best-performing state-of-the-art solutions and applying them to
real-world datasets. We focus on two of the most active application domains
for string dictionaries: Web graph and RDF data. Our results show that our
representations achieve better compression than existing solutions, for
similar query times, and are significantly smaller than any other alternative
that is able to outperform our query times. Overall, our representation, in
its several implementation variants, provides a relevant improvement in
compression relative to previous proposals within very efficient query
times.

We plan to explore the possibilities to extend our ideas to the dynamic
scenario, where insertions and deletions are supported. A direct application 
of our techniques is not feasible in a dynamic environment, since
we use a static decomposition of the collection and 
compression techniques that are also of static nature. However, we believe
that simple adaptations based on the same compression techniques introduced here
would still yield sufficiently compact dynamic dictionaries. Dynamic string
dictionaries in compressed space are useful, for instance, for better
handling large datasets in RDF engines in main memory.

\section{Acknowledgements}

Funded by EU H2020 MSCA RISE grant No 690941 (BIRDS). GN
funded by the Millennium Institute for Foundational Research on Data (IMFD),
and by Fondecyt Grant 1-170048, Conicyt, Chile. NB, ACP and GdB funded by Xunta de Galicia/FEDER-UE grants CSI:
ED431G/01 and GRC:ED431C 2017/58; by MINECO-AEI/FEDER-UE grants
TIN2016-77158-C4-3-R and TIN2016-78011-C4-1-R; and by MICINN grant
RTC-2017-5908-7.

%
\bibliographystyle{ACM-Reference-Format}
\bibliography{paper}


\begin{thebibliography}{21}


\ifx \showCODEN    \undefined \def \showCODEN     #1{\unskip}     \fi
\ifx \showDOI      \undefined \def \showDOI       #1{#1}\fi
\ifx \showISBNx    \undefined \def \showISBNx     #1{\unskip}     \fi
\ifx \showISBNxiii \undefined \def \showISBNxiii  #1{\unskip}     \fi
\ifx \showISSN     \undefined \def \showISSN      #1{\unskip}     \fi
\ifx \showLCCN     \undefined \def \showLCCN      #1{\unskip}     \fi
\ifx \shownote     \undefined \def \shownote      #1{#1}          \fi
\ifx \showarticletitle \undefined \def \showarticletitle #1{#1}   \fi
\ifx \showURL      \undefined \def \showURL       {\relax}        \fi
\providecommand\bibfield[2]{#2}
\providecommand\bibinfo[2]{#2}
\providecommand\natexlab[1]{#1}
\providecommand\showeprint[2][]{arXiv:#2}

\bibitem[\protect\citeauthoryear{Arz and Fischer}{Arz and Fischer}{2018}]%
        {lzstringdict}
\bibfield{author}{\bibinfo{person}{Julian Arz} {and} \bibinfo{person}{Johannes
  Fischer}.} \bibinfo{year}{2018}\natexlab{}.
\newblock \showarticletitle{Lempel---Ziv-78 Compressed String Dictionaries}.
\newblock \bibinfo{journal}{\emph{Algorithmica}} \bibinfo{volume}{80},
  \bibinfo{number}{7} (\bibinfo{date}{July} \bibinfo{year}{2018}),
  \bibinfo{pages}{2012--2047}.
\newblock
\showISSN{0178-4617}
\urldef\tempurl%
\url{https://doi.org/10.1007/s00453-017-0348-7}
\showDOI{\tempurl}


\bibitem[\protect\citeauthoryear{Bille, G{\o}rtz, and Skjoldjensen}{Bille
  et~al\mbox{.}}{2017}]%
        {bille}
\bibfield{author}{\bibinfo{person}{Philip Bille}, \bibinfo{person}{Inge~Li
  G{\o}rtz}, {and} \bibinfo{person}{Frederik~Rye Skjoldjensen}.}
  \bibinfo{year}{2017}\natexlab{}.
\newblock \showarticletitle{Deterministic Indexing for Packed Strings}. In
  \bibinfo{booktitle}{\emph{28th Annual Symposium on Combinatorial Pattern
  Matching, {CPM} 2017, July 4-6, 2017, Warsaw, Poland}}.
  \bibinfo{pages}{6:1--6:11}.
\newblock
\urldef\tempurl%
\url{https://doi.org/10.4230/LIPIcs.CPM.2017.6}
\showDOI{\tempurl}


\bibitem[\protect\citeauthoryear{Boldi and Vigna}{Boldi and Vigna}{2004}]%
        {webgraph}
\bibfield{author}{\bibinfo{person}{Paolo Boldi} {and}
  \bibinfo{person}{Sebastiano Vigna}.} \bibinfo{year}{2004}\natexlab{}.
\newblock \showarticletitle{The Webgraph Framework I: Compression Techniques}.
  In \bibinfo{booktitle}{\emph{Proceedings of the 13th International Conference
  on World Wide Web}} \emph{(\bibinfo{series}{WWW '04})}.
  \bibinfo{publisher}{ACM}, \bibinfo{address}{New York, NY, USA},
  \bibinfo{pages}{595--602}.
\newblock
\showISBNx{1-58113-844-X}
\urldef\tempurl%
\url{https://doi.org/10.1145/988672.988752}
\showDOI{\tempurl}


\bibitem[\protect\citeauthoryear{Brisaboa, Ladra, and Navarro}{Brisaboa
  et~al\mbox{.}}{2013}]%
        {dacs}
\bibfield{author}{\bibinfo{person}{Nieves~R. Brisaboa}, \bibinfo{person}{Susana
  Ladra}, {and} \bibinfo{person}{Gonzalo Navarro}.}
  \bibinfo{year}{2013}\natexlab{}.
\newblock \showarticletitle{DACs: Bringing Direct Access to Variable-length
  Codes}.
\newblock \bibinfo{journal}{\emph{Information Processing and Management}}
  \bibinfo{volume}{49}, \bibinfo{number}{1} (\bibinfo{date}{Jan.}
  \bibinfo{year}{2013}), \bibinfo{pages}{392--404}.
\newblock
\showISSN{0306-4573}
\urldef\tempurl%
\url{https://doi.org/10.1016/j.ipm.2012.08.003}
\showDOI{\tempurl}


\bibitem[\protect\citeauthoryear{Brisaboa, Ladra, and Navarro}{Brisaboa
  et~al\mbox{.}}{2014}]%
        {ktree}
\bibfield{author}{\bibinfo{person}{Nieves~R. Brisaboa}, \bibinfo{person}{Susana
  Ladra}, {and} \bibinfo{person}{Gonzalo Navarro}.}
  \bibinfo{year}{2014}\natexlab{}.
\newblock \showarticletitle{Compact Representation of Web Graphs with Extended
  Functionality}.
\newblock \bibinfo{journal}{\emph{Information Systems}}  \bibinfo{volume}{39}
  (\bibinfo{date}{Jan.} \bibinfo{year}{2014}), \bibinfo{pages}{152--174}.
\newblock
\showISSN{0306-4379}
\urldef\tempurl%
\url{https://doi.org/10.1016/j.is.2013.08.003}
\showDOI{\tempurl}


\bibitem[\protect\citeauthoryear{C.~Hu and C.~Tucker}{C.~Hu and
  C.~Tucker}{1979}]%
        {hutucker}
\bibfield{author}{\bibinfo{person}{T C.~Hu} {and} \bibinfo{person}{A
  C.~Tucker}.} \bibinfo{year}{1979}\natexlab{}.
\newblock \showarticletitle{Optimal Computer Search Trees and Variable-Length
  Alphabetical Codes}.
\newblock \bibinfo{journal}{\emph{Siam Journal on Applied Mathematics -
  SIAMAM}}  \bibinfo{volume}{21} (\bibinfo{date}{Jan.} \bibinfo{year}{1979}).
\newblock
\urldef\tempurl%
\url{https://doi.org/10.1137/0121057}
\showDOI{\tempurl}


\bibitem[\protect\citeauthoryear{Grossi and Ottaviano}{Grossi and
  Ottaviano}{2015}]%
        {pdtries}
\bibfield{author}{\bibinfo{person}{Roberto Grossi} {and}
  \bibinfo{person}{Giuseppe Ottaviano}.} \bibinfo{year}{2015}\natexlab{}.
\newblock \showarticletitle{Fast Compressed Tries Through Path Decompositions}.
\newblock \bibinfo{journal}{\emph{Journal of Experimental Algorithmics}}
  \bibinfo{volume}{19}, Article \bibinfo{articleno}{3.4} (\bibinfo{date}{Jan.}
  \bibinfo{year}{2015}), \bibinfo{numpages}{11}~pages.
\newblock
\showISSN{1084-6654}
\urldef\tempurl%
\url{https://doi.org/10.1145/2656332}
\showDOI{\tempurl}


\bibitem[\protect\citeauthoryear{Huffman}{Huffman}{1952}]%
        {huffman}
\bibfield{author}{\bibinfo{person}{David~A. Huffman}.}
  \bibinfo{year}{1952}\natexlab{}.
\newblock \showarticletitle{A Method for the Construction of Minimum-Redundancy
  Codes}.
\newblock \bibinfo{journal}{\emph{Proceedings of the Institute of Radio
  Engineers}} \bibinfo{volume}{40}, \bibinfo{number}{9} (\bibinfo{date}{Sept.}
  \bibinfo{year}{1952}), \bibinfo{pages}{1098--1101}.
\newblock
\urldef\tempurl%
\url{https://doi.org/10.1109/JRPROC.1952.273898}
\showDOI{\tempurl}


\bibitem[\protect\citeauthoryear{Kanda, Morita, and Fuketa}{Kanda
  et~al\mbox{.}}{2017a}]%
        {kanda1}
\bibfield{author}{\bibinfo{person}{Shunsuke Kanda}, \bibinfo{person}{Kazuhiro
  Morita}, {and} \bibinfo{person}{Masao Fuketa}.}
  \bibinfo{year}{2017}\natexlab{a}.
\newblock \showarticletitle{Compressed double-array tries for string
  dictionaries supporting fast lookup}.
\newblock \bibinfo{journal}{\emph{Knowledge and Information Systems}}
  \bibinfo{volume}{51}, \bibinfo{number}{3} (\bibinfo{year}{2017}),
  \bibinfo{pages}{1023--1042}.
\newblock
\urldef\tempurl%
\url{https://doi.org/10.1007/s10115-016-0999-8}
\showDOI{\tempurl}


\bibitem[\protect\citeauthoryear{Kanda, Morita, and Fuketa}{Kanda
  et~al\mbox{.}}{2017b}]%
        {kanda2}
\bibfield{author}{\bibinfo{person}{Shunsuke Kanda}, \bibinfo{person}{Kazuhiro
  Morita}, {and} \bibinfo{person}{Masao Fuketa}.}
  \bibinfo{year}{2017}\natexlab{b}.
\newblock \showarticletitle{Practical Implementation of Space-Efficient Dynamic
  Keyword Dictionaries}. In \bibinfo{booktitle}{\emph{Proceedings of the 24th
  International Symposium on String Processing and Information Retrieval}}
  \emph{(\bibinfo{series}{SPIRE '17})}, Vol.~\bibinfo{volume}{10508}.
  \bibinfo{publisher}{Springer}, \bibinfo{pages}{221--233}.
\newblock
\urldef\tempurl%
\url{https://doi.org/10.1007/978-3-319-67428-5_19}
\showDOI{\tempurl}


\bibitem[\protect\citeauthoryear{Knuth}{Knuth}{1998}]%
        {knuth1998}
\bibfield{author}{\bibinfo{person}{Donald~E. Knuth}.}
  \bibinfo{year}{1998}\natexlab{}.
\newblock \bibinfo{booktitle}{\emph{{The Art of Computer Programming, volume 3:
  Sorting and searching}}}.
\newblock \bibinfo{publisher}{Addison-Wesley Longman Publishing Co., Inc.},
  \bibinfo{address}{Boston, MA, USA}.
\newblock


\bibitem[\protect\citeauthoryear{Larsson and Moffat}{Larsson and
  Moffat}{1999}]%
        {repair}
\bibfield{author}{\bibinfo{person}{N.~Jesper Larsson} {and}
  \bibinfo{person}{Alistair Moffat}.} \bibinfo{year}{1999}\natexlab{}.
\newblock \showarticletitle{Offline Dictionary-Based Compression}. In
  \bibinfo{booktitle}{\emph{Proceedings of the Conference on Data Compression}}
  \emph{(\bibinfo{series}{DCC '99})}. \bibinfo{publisher}{IEEE Computer
  Society}, \bibinfo{address}{Washington, DC, USA}, \bibinfo{pages}{296--}.
\newblock
\showISBNx{0-7695-0096-X}
\urldef\tempurl%
\url{https://doi.org/10.1109/DCC.1999.755679}
\showDOI{\tempurl}


\bibitem[\protect\citeauthoryear{Manber and W.~Myers}{Manber and
  W.~Myers}{1993}]%
        {suffixArray}
\bibfield{author}{\bibinfo{person}{Udi Manber} {and} \bibinfo{person}{Eugene
  W.~Myers}.} \bibinfo{year}{1993}\natexlab{}.
\newblock \showarticletitle{Suffix Arrays: A New Method for On-Line String
  Searches.}
\newblock \bibinfo{journal}{\emph{SIAM J. Comput.}}  \bibinfo{volume}{22}
  (\bibinfo{date}{Jan.} \bibinfo{year}{1993}), \bibinfo{pages}{935--948}.
\newblock
\urldef\tempurl%
\url{https://doi.org/10.1145/320176.320218}
\showDOI{\tempurl}


\bibitem[\protect\citeauthoryear{Mart\'{\i}nez-Prieto, Brisaboa, C\'{a}novas,
  Claude, and Navarro}{Mart\'{\i}nez-Prieto et~al\mbox{.}}{2016}]%
        {Martinez-Prieto:2016}
\bibfield{author}{\bibinfo{person}{Miguel~A. Mart\'{\i}nez-Prieto},
  \bibinfo{person}{Nieves~R. Brisaboa}, \bibinfo{person}{Rodrigo C\'{a}novas},
  \bibinfo{person}{Francisco Claude}, {and} \bibinfo{person}{Gonzalo Navarro}.}
  \bibinfo{year}{2016}\natexlab{}.
\newblock \showarticletitle{Practical Compressed String Dictionaries}.
\newblock \bibinfo{journal}{\emph{Information Systems}} \bibinfo{volume}{56},
  \bibinfo{number}{C} (\bibinfo{date}{Mar.} \bibinfo{year}{2016}),
  \bibinfo{pages}{73--108}.
\newblock
\showISSN{0306-4379}
\urldef\tempurl%
\url{https://doi.org/10.1016/j.is.2015.08.008}
\showDOI{\tempurl}


\bibitem[\protect\citeauthoryear{Mart\'{\i}nez-Prieto, Fern\'{a}ndez, and
  C\'{a}novas}{Mart\'{\i}nez-Prieto et~al\mbox{.}}{2012}]%
        {rdfdict}
\bibfield{author}{\bibinfo{person}{Miguel~A. Mart\'{\i}nez-Prieto},
  \bibinfo{person}{Javier~D. Fern\'{a}ndez}, {and} \bibinfo{person}{Rodrigo
  C\'{a}novas}.} \bibinfo{year}{2012}\natexlab{}.
\newblock \showarticletitle{Querying RDF Dictionaries in Compressed Space}.
\newblock \bibinfo{journal}{\emph{ACM SIGAPP Applied Computing Review}}
  \bibinfo{volume}{12}, \bibinfo{number}{2} (\bibinfo{date}{June}
  \bibinfo{year}{2012}), \bibinfo{pages}{64--77}.
\newblock
\showISSN{1559-6915}
\urldef\tempurl%
\url{https://doi.org/10.1145/2340416.2340422}
\showDOI{\tempurl}


\bibitem[\protect\citeauthoryear{Navarro}{Navarro}{2016}]%
        {cds}
\bibfield{author}{\bibinfo{person}{Gonzalo Navarro}.}
  \bibinfo{year}{2016}\natexlab{}.
\newblock \bibinfo{booktitle}{\emph{Compact Data Structures: A Practical
  Approach} (\bibinfo{edition}{1st} ed.)}.
\newblock \bibinfo{publisher}{Cambridge University Press},
  \bibinfo{address}{New York, NY, USA}.
\newblock
\showISBNx{9781107152380}
\urldef\tempurl%
\url{https://doi.org/10.1017/CBO9781316588284}
\showDOI{\tempurl}


\bibitem[\protect\citeauthoryear{Neumann and Weikum}{Neumann and
  Weikum}{2010}]%
        {rdfx}
\bibfield{author}{\bibinfo{person}{Thomas Neumann} {and}
  \bibinfo{person}{Gerhard Weikum}.} \bibinfo{year}{2010}\natexlab{}.
\newblock \showarticletitle{The RDF-3X Engine for Scalable Management of RDF
  Data}.
\newblock \bibinfo{journal}{\emph{The VLDB Journal}} \bibinfo{volume}{19},
  \bibinfo{number}{1} (\bibinfo{date}{Feb.} \bibinfo{year}{2010}),
  \bibinfo{pages}{91--113}.
\newblock
\showISSN{1066-8888}
\urldef\tempurl%
\url{https://doi.org/10.1007/s00778-009-0165-y}
\showDOI{\tempurl}


\bibitem[\protect\citeauthoryear{Okanohara and Sadakane}{Okanohara and
  Sadakane}{2007}]%
        {sdarray}
\bibfield{author}{\bibinfo{person}{Daisuke Okanohara} {and}
  \bibinfo{person}{Kunihiko Sadakane}.} \bibinfo{year}{2007}\natexlab{}.
\newblock \showarticletitle{Practical Entropy-compressed Rank/Select
  Dictionary}. In \bibinfo{booktitle}{\emph{Proceedings of the Meeting on
  Algorithm Engineering \& Experiments}}. \bibinfo{publisher}{Society for
  Industrial and Applied Mathematics}, \bibinfo{address}{Philadelphia, PA,
  USA}, \bibinfo{pages}{60--70}.
\newblock
\urldef\tempurl%
\url{https://doi.org/10.1137/1.9781611972870.6}
\showDOI{\tempurl}


\bibitem[\protect\citeauthoryear{Raman, Raman, and Rao}{Raman
  et~al\mbox{.}}{2002}]%
        {rrr}
\bibfield{author}{\bibinfo{person}{Rajeev Raman}, \bibinfo{person}{Venkatesh
  Raman}, {and} \bibinfo{person}{S.~Srinivasa Rao}.}
  \bibinfo{year}{2002}\natexlab{}.
\newblock \showarticletitle{Succinct Indexable Dictionaries with Applications
  to Encoding K-ary Trees and Multisets}. In
  \bibinfo{booktitle}{\emph{Proceedings of the Thirteenth Annual ACM-SIAM
  Symposium on Discrete Algorithms}} \emph{(\bibinfo{series}{SODA '02})}.
  \bibinfo{publisher}{Society for Industrial and Applied Mathematics},
  \bibinfo{address}{Philadelphia, PA, USA}, \bibinfo{pages}{233--242}.
\newblock
\showISBNx{0-89871-513-X}
\urldef\tempurl%
\url{https://doi.org/10.1145/1290672.1290680}
\showDOI{\tempurl}


\bibitem[\protect\citeauthoryear{Tsuruta, K{\"{o}}ppl, Kanda, Nakashima,
  Inenaga, Bannai, and Takeda}{Tsuruta et~al\mbox{.}}{2019}]%
        {dpct}
\bibfield{author}{\bibinfo{person}{Kazuya Tsuruta}, \bibinfo{person}{Dominik
  K{\"{o}}ppl}, \bibinfo{person}{Shunsuke Kanda}, \bibinfo{person}{Yuto
  Nakashima}, \bibinfo{person}{Shunsuke Inenaga}, \bibinfo{person}{Hideo
  Bannai}, {and} \bibinfo{person}{Masayuki Takeda}.}
  \bibinfo{year}{2019}\natexlab{}.
\newblock \showarticletitle{Dynamic Packed Compact Tries Revisited}.
\newblock \bibinfo{journal}{\emph{CoRR}}  \bibinfo{volume}{abs/1904.07467}
  (\bibinfo{year}{2019}).
\newblock
\showeprint{1904.07467}


\bibitem[\protect\citeauthoryear{Williams and Zobel}{Williams and
  Zobel}{1999}]%
        {vbyte}
\bibfield{author}{\bibinfo{person}{Hugh~E. Williams} {and}
  \bibinfo{person}{Justin Zobel}.} \bibinfo{year}{1999}\natexlab{}.
\newblock \showarticletitle{{Compressing Integers for Fast File Access}}.
\newblock \bibinfo{journal}{\emph{Comput. J.}} \bibinfo{volume}{42},
  \bibinfo{number}{3} (\bibinfo{date}{Jan.} \bibinfo{year}{1999}),
  \bibinfo{pages}{193--201}.
\newblock
\showISSN{0010-4620}
\urldef\tempurl%
\url{https://doi.org/10.1093/comjnl/42.3.193}
\showDOI{\tempurl}


\end{thebibliography}

%
%
%
%
%
%
%
%
%

\end{document}